\documentclass[aps,prb,twocolumn,twoside,floatfix,amsmath,showpacs]{revtex4}

\usepackage{graphicx}
\usepackage{natbib}
\begin{document}

\newcommand{\cerusi}{CeRu$_2$Si$_2$}
\newcommand{\ybrhsi}{YbRh$_2$Si$_2$}
\newcommand{\urusi}{URu$_2$Si$_2$}
\newcommand{\cecoin}{CeCoIn$_5$}
\newcommand{\upt}{UPt$_3$}
\newcommand{\thcrsi}{ThCr$_2$Si$_2$}
\newcommand{\threetwoseven}{Sr$_3$Ru$_2$O$_7$}

\newcommand{\sovert}[1]{\ensuremath{#1\,\mu\textnormal{V K}^{-2}}}
\newcommand{\mJmolK}[1]{\ensuremath{#1\,\textnormal{mJ mol}^{-1} \textnormal{K}^{-2}}}
\newcommand{\lorenzunits}{\ensuremath{\,\textnormal{W \Omega K}^{-2}}}

\newcommand{\ie}{{\em i.e.}}

\title{Thermoelectric transport across the metamagnetic transition of \cerusi{}}

\author{Heike Pfau}
\author{Ramzy Daou}
\author{Manuel Brando}
\author{Frank Steglich}
\affiliation{Max Planck Institute for Chemical Physics of Solids, N\"{o}thnitzer
Str. 40, 01187 Dresden, Germany}

\date{\today}


\begin{abstract}
We have measured the thermopower across the metamagnetic transition of the heavy fermion compound \cerusi{} at temperatures down to 0.1\,K and magnetic fields up to 11.5\,T. We find a large negative enhancement of the thermopower on crossing the metamagnetic field, as well as a sudden change in slope. We argue that this is consistent with the Zeeman-driven deformation of the Fermi surface through a topological transition. The field dependence of the thermopower highlights the discrepancy between thermodynamic and transport properties across the metamagnetic transition.
\end{abstract}

\maketitle

\section{Introduction}

In heavy fermion systems, the weak hybridisation between nearly localized $f$-electrons and the mobile conduction electrons leads to a Fermi liquid ground state with narrow bands and quasiparticles with strongly enhanced effective electronic masses. The heavy fermion bandwidth can become comparable to the Zeeman energy in accessible magnetic fields. The Fermi surface can therefore be significantly deformed as the bands are spin-split by an applied magnetic field, and topological (or Lifshitz~\cite{Lifshitz1960}) transitions are possible. These transitions break no symmetry and appear thermodynamically as crossovers at any finite temperature~\cite{Blanter1994}. They are, however, true quantum phase transitions in the limit of low temperature and are therefore surrounded by a regime of quantum critical fluctuations. There has been renewed interest recently in the possible connection between Lifshitz transitions and quantum critical behaviour in heavy fermion and strongly correlated materials as an alternative to scenarios which involve competition with the Kondo mechanism~\cite{Hackl2011,Altarawneh2011,Daou2006a,Gorkov2006,Kotegawa2011,Imada2010,Malone2011,Yelland2011,Rost2011}. 

In this context the heavy fermion paramagnet \cerusi{} has attracted much interest. For a good review of the many thermodynamic, transport and spectroscopic experiments on \cerusi{}, see Ref.~\onlinecite{Flouquet2002}. When a magnetic field of $\sim 7.8$\,T is applied along the crystalline $c$-axis, all thermodynamic properties show a large anomaly, the most obvious of which are a rapid non-linear rise in the magnetisation and a pronounced peak in the electronic specific heat, $\gamma$. The thermodynamics of this {\em metamagnetic transition} (MMT) are consistent with proximity to a quantum critical end point~\cite{Weickert2010,Millis2002}. In the low temperature limit, however, the material exhibits Fermi liquid like properties at all magnetic fields and the MMT is always a crossover. The phase diagram is uncomplicated by ordered phases such as magnetism or superconductivity. Meanwhile, low temperature Hall effect measurements show a very small but distinct anomaly that appears to be sharper than the thermodynamic signatures of the MMT~\cite{Daou2006a}, but occurs at the same critical magnetic field. This was interpreted as a sign of a topological transition of the Fermi surface, driven by enhanced Zeeman splitting of the heavy fermion bands. Indeed, spin-splitting and considerable spin-dependence of the quasiparticle masses was directly detectable via de Haas-van Alphen (dHvA) measurements in some of the lighter bands~\cite{Daou2006,Takashita1996}. 

How then, can we explain relatively sharp features in transport apparently coupled to smooth behaviour in the thermodynamics? Why do they occur at the same magnetic field? To address these questions we have carried out a high resolution study of the thermopower and thermal conductivity of \cerusi{} in magnetic fields up to 11.5\,T and to temperatures down to 0.1\,K. The thermopower is affected by both the scattering processes and thermodynamic properties of the electron fluid, and so is an ideal probe to investigate the connection between the two. It is also predicted to acquire a singular part close to a Lifshitz transition~\cite{Blanter1994}.

In this paper we show that the low temperature behaviour of the thermal transport properties of \cerusi{} are also compatible with a Lifshitz transition model. We find multiple sign changes in the field dependent thermopower, as well as a large negative peak at the MMT that persists in the low temperature limit. There is a sharp kink in the thermopower precisely at the metamagnetic field, similar to that seen in the Hall effect. Using simple models we can reproduce either the sharp features in transport, or the peak in $\gamma$, but not both simultaneously. We conclude that single-particle models with rigid band shifts are alone insufficient to describe the MMT and must be extended.

\section{Experimental}

\subsection{Methods}
The sample of \cerusi{} used here was grown by F.S.~Tautz in Cambridge. The residual resistivity, $\rho_0$, was 1.1\,$\mu\Omega$cm, compared to 0.4\,$\mu\Omega$cm for a sample from the same batch measured in Ref.~\onlinecite{Daou2006a}. Thermal transport properties were measured on a single crystal of dimensions 3.5$\times$0.28$\times$0.08\,mm$^3$ ($a\times b\times c$) using the standard one-heater, two-thermometer technique. Low resistance contacts were made by soldering to the sample. A systematic uncertainty of about 10\% that reflects the size of the contacts is ignored here as it acts as a scaling factor only. Magnetic field was applied along the crystalline $c$-axis, while thermal and electrical currents were applied in the $ab$-plane. Superconducting NbTi filaments were used as the reference leads for the thermoelectric voltage measurement, resulting in background-free data over the whole field and temperature range studied. These filaments also permitted {\em in-situ} four-point d.c. resistivity measurements. The current and magnetic field were reversed to check for thermoelectric contributions to the d.c. resistivity, but these were found to be negligible for the currents employed (10\,$<I<$\,150\,$\mu$A). This was also a useful check to confirm that transverse transport properties (\ie{} Hall, thermal Hall and Nernst effects) made no detectable contribution to the measurements. Temperature sweeps were performed at several fixed magnetic fields up to 11.5\,T. Additionally, magnetic field sweeps were perfomed at several fixed temperatures. In order to avoid the high thermal resistance of superconducting contacts used in the experimental stage, the minimum field we applied was 0.2\,T. 

\subsection{Results}

\begin{figure}
  \begin{center}
    \includegraphics[width=\linewidth]{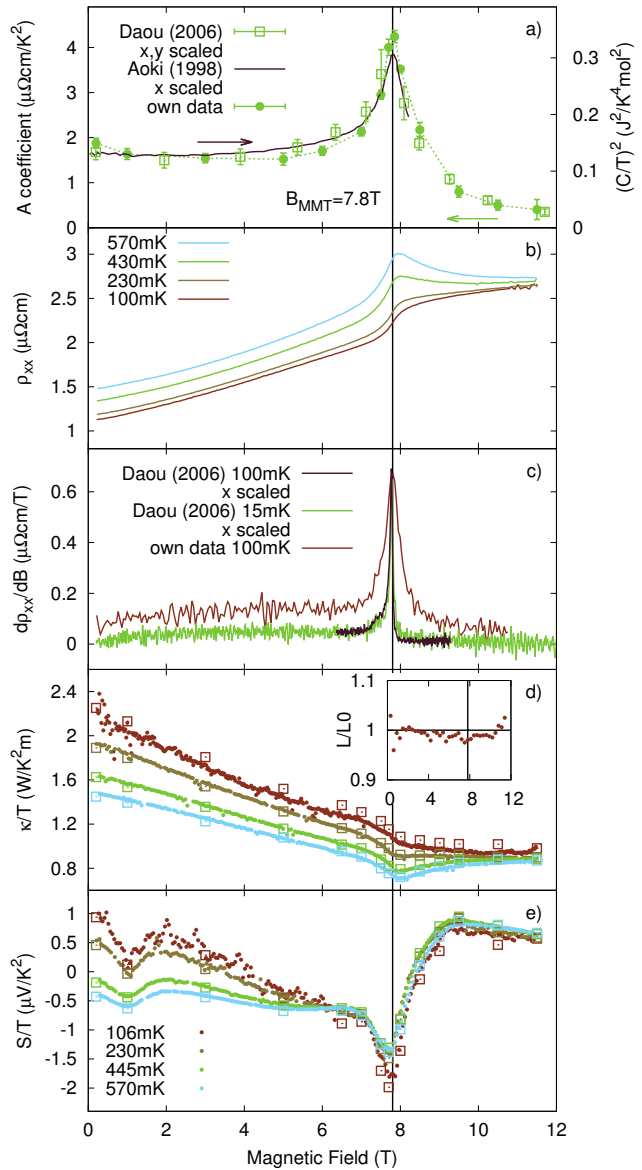}
  \end{center}
  \caption{(Color online) Low temperature transport properties of \cerusi{} as a function of magnetic field, $B$. a) The $A$-coefficient of the resistivity, $\rho$, peaks at $B_{MMT}=7.80 \pm 0.05T$. For comparison, similar data is included for the sample of Ref.~\onlinecite{Daou2006a} and also published specific heat data on another sample~\cite{Aoki1998}. b) A peak in $\rho(B)$ appears at higher temperatures (lighter curves). The resistivity measured at 100\,mK is close to the residual resistivity $\rho_0$, however, and shows no such peak at the MMT. c) The derivative of $\rho_0$ with respect to magnetic field, compared to the sample of Ref.~\onlinecite{Daou2006a}. The data of Ref.~\onlinecite{Daou2006a} presented in a) and c) has been slightly scaled so that $B_{MMT}$ matches, accounting for a misalignment of the sample. The MMT appears much broader in $\rho$ of the present sample. d) The thermal conductivity. Field sweeps (filled circles) and temperature sweeps (open squares) are in excellent agreement. The inset shows the Lorenz ratio at 100\,mK. e) The thermopower divided by temperature shows multiple sign changes and a strong negative peak near the MMT. There is a sharp kink in the thermopower precisely at $B_{MMT}$ which is more clearly seen in Fig.\ref{fig:qratio}. An additional peak at 1\,T becomes sharper as the temperature is reduced.}
  \label{fig:overview}
\end{figure}

\subsubsection{Magnetoresistance}

We begin by comparing the magnetoresistance data to previously available data from Ref.~\onlinecite{Daou2006a}. As in previous studies, at low temperatures the resistivity ($\rho$) is Fermi-liquid like, {\em i.e.} $\rho=\rho_0 +AT^2$, at all fields. Figures~\ref{fig:overview}a) and b) show $A$ and $\rho$ extracted from temperature and field sweeps respectively. $A$ is proportional to the square of the electronic specific heat coefficient, $\gamma$, according to the Kadowaki-Woods relationship~\cite{Kadowaki1986} and is hence a thermodynamic property. This relationship has been seen to hold well in other samples~\cite{Kambe1996,Aoki2011}. Clearly the width of the peak in $A$ does not have a strong sample dependence. Looking at $d\rho/dB$ in Fig.~\ref{fig:overview}c), however, we can see that the peak at the MMT is much narrower for the sample in Ref.~\onlinecite{Daou2006a}. Small differences in sample quality affect $\rho_0$, which arises only from electron-impurity scattering, much more than thermodynamic properties.

Examining the data from Ref.~\onlinecite{Daou2006a} more carefully, we see that $d\rho/dB$ peaks at precisely the same field as that where the small negative dip in the Hall resistivity occurs. In the current data, the peak in $d\rho/dB$ is at $B_{MMT}=7.80\pm0.05$\,T. This coincides with the peak in $A$ extracted from the temperature dependence of the resistivity at fixed fields, within the resolution of the measurement.

Other than the ordinary magnetoresistance arising from cyclotron motion of the electrons, variation of $\rho_0$ with magnetic field can arise either from changes in the shape of the Fermi surface (orbital contributions) or from modification of the impurity potential, perhaps due to enhanced zero-temperature fluctuations~\cite{Miyake2002}. Calculations based on this second mechanism in the vicinity of a ferromagnetic quantum critical point show that the fluctuations generate a peak in $\rho_0$ at the critical field. We know that the ferromagnetic fluctuations in \cerusi{} are also peaked around the MMT~\cite{Sato2004}. $\rho_0$ shows a monotonic increase, however, suggesting that the orbital mechanism is more important.

\subsubsection{Thermal Conductivity}

The thermal conductivity, $\kappa$, is shown in Figure~\ref{fig:overview}d. When only electronic transport is involved, it is directly related to the electrical conductivity, $\sigma$, via the Wiedemann-Franz law (WFL), which is written as $\kappa/\sigma T = L_0$ where $L_0=\frac{\pi^2 k_B^2}{3 e^2}$ is Sommerfeld's constant. This ratio holds to within 5\% at all fields at 100\,mK, as we might expect in the Fermi liquid regime (see inset to Fig.~\ref{fig:overview}d). Previous studies down to 0.64\,K were not able to test the validity of the WFL in the low temperature limit~\cite{Amato1989,Sera1997}.

\subsubsection{Thermoelectric power}

The thermopower, $S$, of \cerusi{} has a complex temperature dependence reflecting the compensated multi-band electronic structure~\cite{Amato1989}. There are multiple sign changes in the temperature range below 10\,K. At low temperature and low fields $S/T$ reaches a value of \sovert{0.7}, considerably smaller than that reported by Ref.~\onlinecite{Amato1989}. This difference may be the result of the small 0.2\,T field that was the lowest field that we applied, or it may reflect the dependence of the thermopower in a compensated multiband metal on sample quality.

Figure \ref{fig:overview}e) shows that $S(B)/T$ changes sign twice at low temperature as the magnetic field increases. There is a broad negative peak in the thermopower, apparently at a field slightly below $B_{MMT}$. There is an additional peak at 1\,T that becomes more sharply defined at low temperatures which is not associated with any known feature in other transport or thermodynamic properties. This peak is not resolved in measurements at 1.5\,K~\cite{BelThesis2004}.

\begin{figure}
  \begin{center}
    \includegraphics[angle=-90,width=\linewidth]{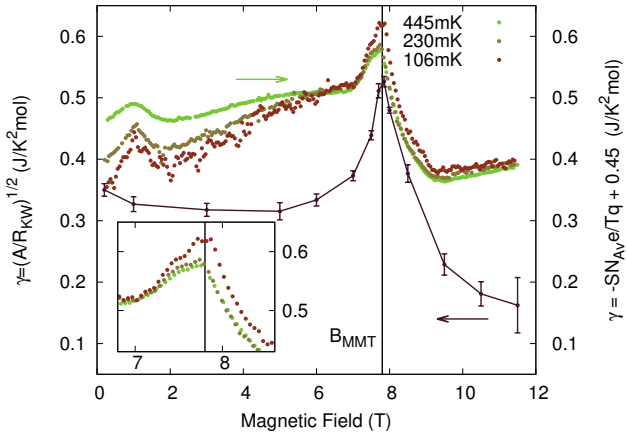}
  \end{center}
  \caption{(Color online) Comparing thermopower to thermodynamics. We calculate the electronic specific heat coefficient, $\gamma$, in two ways. On the left scale we plot $\gamma$ obtained from the Kadowaki-Woods ratio, by using the $A$-coefficient of the electrical resistivity in the formula $A/\gamma^2=R_{KW}$. On the right scale we use the $q$-ratio (see text) to scale the thermopower data into units of $\gamma$. We also invert the sign and offset it by 0.45\,J\,mol$^{-1}$\,K$^{-2}$ to match the left scale at zero field. An offset is justified by the presence of other bands, while the sign inversion is appropriate for heavy fermion materials. Clearly, the structure in the thermopower near the MMT is more complex than the thermodynamic signature. The inset shows an enlargement at $B_{MMT}$ showing the sharp kink in the thermopower-derived data which persists to higher temperatures.}
  \label{fig:qratio}
\end{figure}

Precisely at $B_{MMT}$, there is a sharp kink in the thermopower. $S/T(B)$ changes slope suddenly, within the resolution of our measurement, about 0.05\,T. This feature is more clearly seen in Fig.~\ref{fig:qratio}. This kink is resolved up to the highest temperature that we measure (0.57\,K). This temperature dependence is similar to that of the narrow feature in the Hall effect at $B_{MMT}$, which is clear and essentially unchanged below 0.5\,K \cite{Daou2006a}.

\paragraph{Sign of the thermopower}
The sign of the thermopower is often used as an indicator of the sign of the dominant carriers. An electron band would make a negative contribution to the thermopower, and a hole band a positive contribution. In the case of heavy fermion materials, this general trend is inverted. Ce-based heavy fermion materials typically have a positive thermopower in the low temperature limit~\cite{Behnia2004}, even though they usually have electron-like Fermi surfaces. This sign inversion is an expected consequence of the heavy fermion state~\cite{Miyake2005,Zlatic2007}.

In the particular case of \cerusi{}, the positive Hall effect suggests that hole-like carriers dominate the transport at all fields. The positive $S/T$ at $B=0$ implies conversely that it is electron-like carriers that dominate. If we assume that the large hole-like sheet of the Fermi surface with the most enhanced effective mass dominates the charge transport, we would expect a {\em negative} thermopower. In a multiband system, however, the sum of contributions to the thermopower is weighted by the band conductivity, and $S$ can acquire either sign depending on the details of the bandstructure. The positive `background' $S/T$ at low and high fields would be a result of this. We could then interpret the large negative peak in $S(B)/T$ as an enhancement of the contribution of a hole-like surface with significant $f$-character that is tuned with magnetic field, with a maximum effect near $B_{MMT}$. The natural candidate for this is the heavy hole band that has been observed by dHvA.

\paragraph{q-ratio}
An empirical relationship between $S/T$ as $T \rightarrow 0$ and $\gamma$ has been established in heavy fermion systems~\cite{Behnia2004}. This relationship is quantified by the $q$-ratio, where $q=\frac{S}{T}\frac{N_{Av}e}{\gamma} \approx \pm 1$. $N_{Av}$ is Avogadro's number. It is based on a series of approximations; in particular that the derivative of the electrical conductivity, $\sigma$, in the Mott formula, $S/T = L_0e(\partial \ln \sigma / \partial \varepsilon)_{\varepsilon_F}$, can be written as $(\partial \ln \tau(\varepsilon)/\partial\varepsilon + \partial \ln N(\varepsilon)/\partial\varepsilon)_{\varepsilon_F}\approx 1/\varepsilon_F$. $N(\varepsilon)$ is the electronic density of states and $\tau$ is the scattering lifetime.

In Fig.~\ref{fig:qratio} we show $\gamma$ derived from the $A$-coefficient ($\gamma_A$) via the Kadowaki-Woods relationship and compare it to that derived from $S/T$ via the $q$-ratio ($\gamma_S$). The enhancement of $\gamma_A$ at the MMT compared to the zero field value is similar in magnitude to the enhancement of $\gamma_S$. 
The detailed field dependence is very different, however. The peak at 1\,T is not correlated with any feature seen in other thermodynamic or transport properties.
The kink in $S/T$ at $B_{MMT}$ corresponds to the location of the peak in $A$. Both features have little temperature dependence below 0.5\,K~\cite{Weickert2010}. This is seen more clearly in the inset to Fig.~\ref{fig:qratio}. 


The successive approximations involved in estimating the $q$-ratio impose severe limitations on the validity of this test, in particular if either $N(\varepsilon)$ or $\tau$ have a strong dependence on energy. In a narrow band system the magnetic field may then cause strong variations in either quantity as the Zeeman energy increases. At a topological transition of the Fermi surface in three dimensions
either $(\partial \ln N(\varepsilon)/\partial\varepsilon)_{\varepsilon_F}$ or $(\partial \ln \tau(\varepsilon)/\partial\varepsilon)_{\varepsilon_F}$ diverges when $N(\varepsilon)$ does not. 
Sharp features in $S/T$ without thermodynamic counterparts are then possible. The peak at 1\,T may be one such feature. One way of distinguishing whether it is the scattering or the thermoydnamic term that is important might be to measure the Nernst effect, which is (in some limit) only dependent on $(\partial \tau (\varepsilon) / \partial\varepsilon)_{\varepsilon_F}$~\cite{Behnia2009}. 

\section{Rigid band shift models}

\begin{figure}
  \begin{center}
  \includegraphics[angle=0,width=\linewidth]{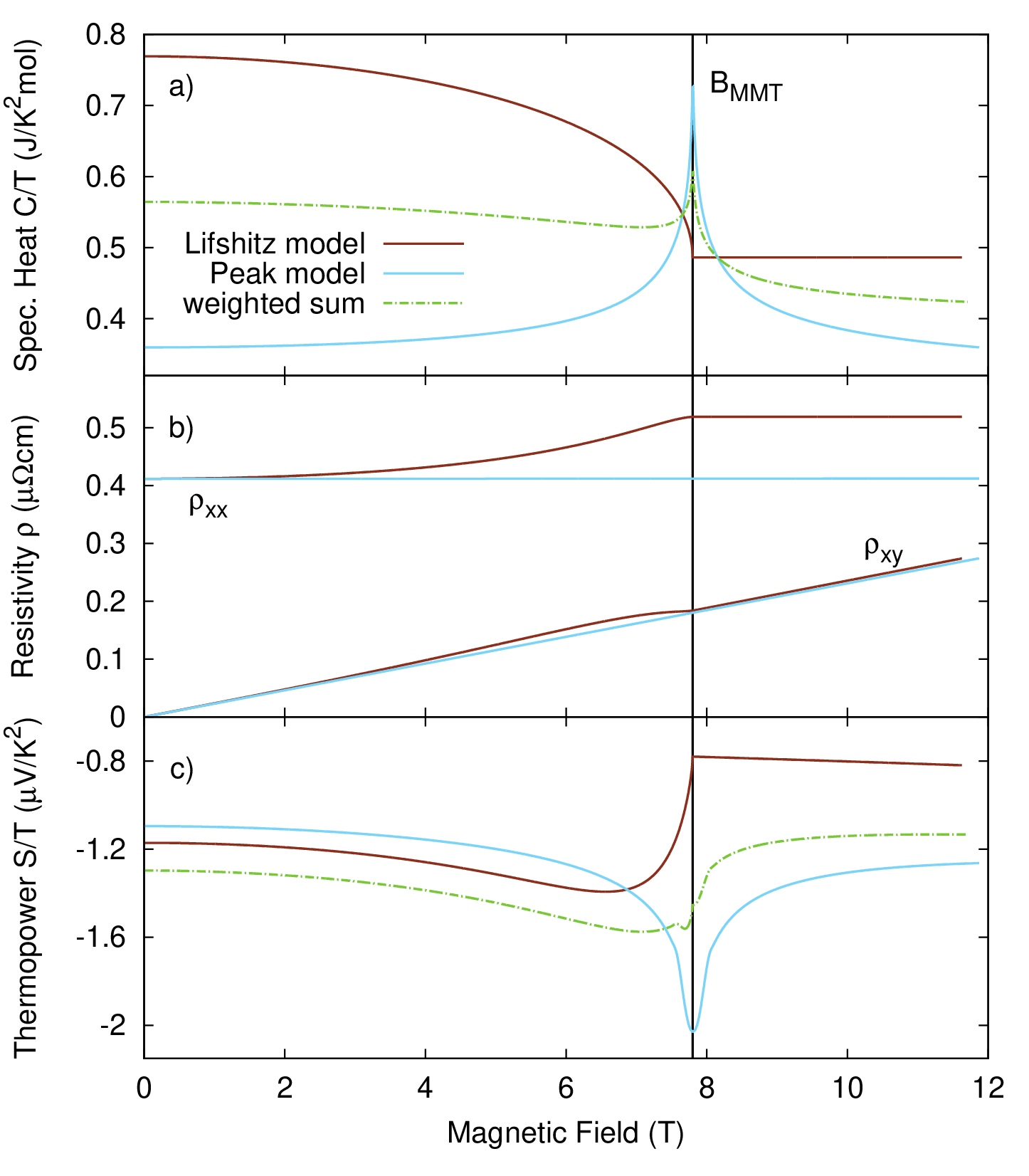}
  \end{center}
    \caption{(Color online) Model calculations of transport and thermodynamic properties at zero temperature. In the `Lifshitz' model (dark red), the spherical Fermi surface is spin-split until one band becomes completely unoccupied at $B_{MMT}$. The total number of occupied states is conserved. In the `peak' model (light blue), the Fermi energy passes through a peak in the density of states at $B_{MMT}$. A combination of these models (see text for details, dashed green) can qualitatively reproduce key features of both transport and thermodynamic properties. a) The specific heat, $C/T$. The observed specific heat can be well reproduced by the peak model. b) The resistivity and Hall effect only have sharp features in the Lifshitz model calculation. c) The thermopower of the Lifshitz model has an asymmetric structure and a sharp kink at $B_{MMT}$.}
  \label{fig:model}
\end{figure}

We consider now two simple model calculations that help us to understand the generic features of transport and thermodynamics. Both models use a single, spin-split spherical band with parameters drawn from dHvA measurements. In the `Lifshitz' model (also discussed in Ref.~\onlinecite{Daou2006a}), the splitting is so strong that one spin-subband becomes completely depopulated at $B_{MMT}$. In the `peak' model, $\varepsilon_F$ is driven through a sharp peak in the density of states at $B_{MMT}$ (as suggested in Ref.~\onlinecite{Aoki1998}). We extend the transport calculation of Ref.~\onlinecite{Daou2006a} to include the thermopower by applying the Mott formula, and we also calculate the electronic specific heat. These calculations are shown in Figure~\ref{fig:model}. As we can see, $C/T$ can be well described by the peak model. However, there are no sharp features in the transport propeties. The Lifshitz model, meanwhile, can reproduce well the sharp features seen in the Hall effect. The calculated $S(B)/T$ also has an asymmetric shape and a sharp kink, qualitatively similar to the data. It is difficult to otherwise reproduce the sharp structure seen in the transport simply by invoking sharp features in $N(\varepsilon)$; if this were the case one would expect to see features of similar width in magnetic field in both thermodynamic and transport properties. However, the Lifshitz model cannot reasonably produce a peak in $C/T$.

Interestingly, a superposition of these two models would be able to account for both the transport and thermodynamic anomalies. Such a combined model would have a peak in the density of states arising from a non-vanishing band coincident with the edge of another band. This idea is illustrated in Figure~\ref{fig:model}a, where $C/T$ is shown arising from two such contributions (in this case we take the average $C/T$). In the same vein, if we construct $S(B)/T$ from two contributions, similar in shape to those calculated for the Lifshitz and peak models, we arrive at a curve that bears a qualitative resemblence to the one observed (see Fig.~\ref{fig:model}c, which shows a weighted sum of the two model calculations to illustrate this. A weighted sum is appropriate for the thermopower since $S=\sum_i\frac{S_i\sigma_i}{\sigma}$). While we can envisage a bandstructure that contains both a peak and a band egde, if it is to be more than simply serendipitous that they are both traversed at the same magnetic field then a further physical connection is required. There must be some reason for the peak to be pinned to the band edge. We note that similar simultaneous signatures in transport and thermodynamic properties have been recently observed in another material~\cite{Yelland2011}, making a generic mechanism more likely than coincidence.

One possibility is that the `peak' contribution arises from fluctuations that originate in the metamagnetic fluctuations between states of low and high magnetisation. These may also be viewed as topological fluctuations between the state with an extra Fermi surface pocket, and the one where it has vanished. The Lifshitz transition is a kind of quantum phase transition, and therefore we expect a regime of fluctuations to surround it. While one band vanishes, these fluctuations would renormalise the electronic density of states in every band, most strongly at $B_{MMT}$.

\begin{figure}
  \begin{center}
    \includegraphics[angle=0,width=\linewidth]{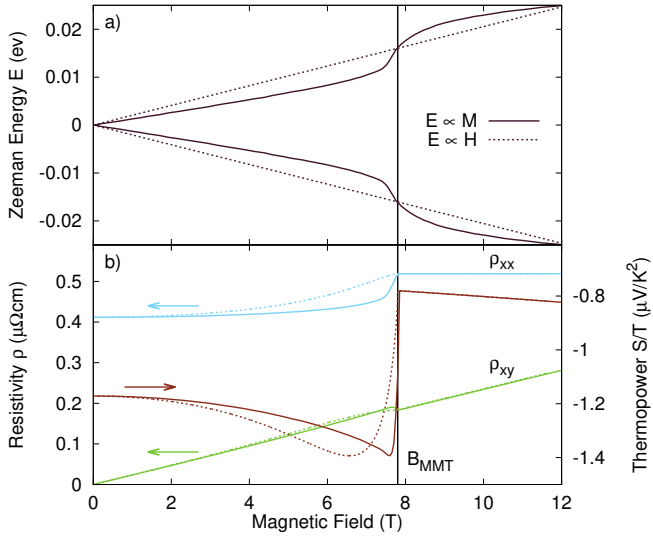}
  \end{center}
  \caption{(Color online) Lifshitz model calculation with experimental field dependence. The Zeeman energy, $E$, is plotted in the top panel. Dashed lines show the results for $E\propto B$ and solid lines for $E\propto M$. The peak in the measured thermopower (dark red, right axis) is asymmetric with respect to the critical field $B_{MMT}$, with a greater apparent weight for $B<B_{MMT}$. If the Zeeman splitting is instead taken to be proportional to the experimental magnetisation (from Ref.~\onlinecite{Flouquet2002}, dashed lines) the features in the calculation appear much narrower.}
  \label{fig:originalmodel}
\end{figure}

While there has been some recent interest in quantum topological transitions in metamagnetic systems~\cite{Imada2010,Rost2011}, a full theoretical treatment is still lacking and the ideas presented here remain quite speculative. It is clear that non-interacting models with rigid band shifts are insufficient to explain the data we obtain and it may therefore be interesting to explore field dependent renormalized band calculations, as have been performed successfully for \ybrhsi{}~\cite{Zwicknagel2011}, where the field dependence of the heavy fermion bandstructure is accounted for. 
As an illustration of the possible effect of strongly field dependent parameters, we show again in Figure~\ref{fig:originalmodel} the Lifshitz model calculation, repeated with a Zeeman energy taken to be proportional to the experimental magnetisation. The transition has the same general features, but is compressed into a narrower field range.

\section{Summary}

We have presented high resolution measurements of the thermal transport properties of \cerusi{} across the MMT. The thermopower, like the Hall effect, shows a sharp feature at the MMT which is difficult to explain without a singular contribution to the density of states. However, the thermodynamic properties at the MMT are not divergent. A topological transition of the Fermi surface is one way in which transport anomalies can be explained. We have tried to show how the thermodynamics might also be accounted for if quantum fluctuations are considered, but we await a rigorous theoretical treatment of the interacting Lifshitz point.

We acknowledge helpful discussions with V. Zlati\'{c}, M. Vojta and C. Geibel.

\end{document}